\documentclass[sigconf]{acmart}

\usepackage{textcomp}
\usepackage{xcolor}
\usepackage{bm}
\usepackage{graphicx}
\usepackage{balance}
\usepackage{textcomp}
\usepackage[export]{adjustbox}
\usepackage{booktabs}
\usepackage{tabularx}
\usepackage{footnote}
\usepackage{array}
\newcolumntype{P}[1]{>{\centering\arraybackslash}p{#1}}
\newcolumntype{M}[1]{>{\centering\arraybackslash}m{#1}}
\usepackage{multirow}
\usepackage{url}
\usepackage{algorithm}
\usepackage[noend]{algpseudocode}
\usepackage{dsfont}
\usepackage[bottom]{footmisc}
\usepackage[inline]{enumitem}
\usepackage{subcaption}

\newtheorem{definition}{Definition}
\newtheorem{theorem}{Theorem}
\DeclareMathOperator{\real}{real}
\DeclareMathOperator{\synth}{synth}

\newcommand{\eat}[1]{}
\newcommand{\para}[1]{\smallskip\noindent\textit{#1.}}

\copyrightyear{2021} 
\acmYear{2021} 
\setcopyright{acmcopyright}
\acmConference[SSTD '21]{17th International Symposium on Spatial and Temporal Databases}{August 23--25, 2021}{virtual, USA}
\acmBooktitle{17th International Symposium on Spatial and Temporal Databases (SSTD '21), August 23--25, 2021, virtual, USA}
\acmPrice{15.00}
\acmDOI{10.1145/3469830.3470893}
\acmISBN{978-1-4503-8425-4/21/08}

\begin{document}

\title{Privacy-Preserving Synthetic Location Data in the Real World}

\author{Teddy Cunningham}
\affiliation{
  \institution{University of Warwick}
  \city{Coventry}
  \country{United Kingdom}
  \postcode{CV4 7AL}
}
\email{teddy.cunningham@warwick.ac.uk}

\author{Graham Cormode}
\affiliation{
  \institution{University of Warwick}
  \city{Coventry}
  \country{United Kingdom}
  \postcode{CV4 7AL}
}
\email{g.cormode@warwick.ac.uk}

\author{Hakan Ferhatosmanoglu}
\affiliation{
  \institution{University of Warwick}
  \city{Coventry}
  \country{United Kingdom}
  \postcode{CV4 7AL}
}
\email{hakan.f@warwick.ac.uk}

\renewcommand{\shortauthors}{Cunningham, et al.}

\begin{abstract}
Sharing sensitive data is vital in enabling many modern data analysis and machine learning tasks. 
However, current methods for data release are insufficiently accurate or granular to provide meaningful utility, and they carry a high risk of deanonymization or membership inference attacks.
In this paper, we propose a differentially private synthetic data generation solution with a focus on the compelling domain of location data.
We present two methods with high practical utility for generating synthetic location data from real locations, both of which protect the existence and true location of each individual in the original dataset. 
Our first, partitioning-based approach introduces a novel method for privately generating point data using kernel density estimation, in addition to employing private adaptations of classic statistical techniques, such as clustering, for private partitioning.
Our second, network-based approach incorporates public geographic information, such as the road network of a city, to constrain the bounds of synthetic data points and hence improve the accuracy of the synthetic data.
Both methods satisfy the requirements of differential privacy, while also enabling accurate generation of synthetic data that aims to preserve the distribution of the real locations.
We conduct experiments using three large-scale location datasets to show that the proposed solutions generate synthetic location data with high utility and strong similarity to the real datasets. 
We highlight some practical applications for our work by applying our synthetic data to a range of location analytics queries, and we demonstrate that our synthetic data produces near-identical answers to the same queries compared to when real data is used. 
Our results show that the proposed approaches are practical solutions for sharing and analyzing sensitive location data privately.
\vspace{0.8cm}
\end{abstract}

\ccsdesc[500]{Information systems~Spatial-temporal systems}
\ccsdesc[500]{Security and privacy~Privacy protections}
\ccsdesc[300]{Information systems~Data analytics}

\keywords{Differential Privacy, Location Data Sharing, Synthetic Data}
\maketitle

\section{Introduction}
\label{s:intro}
People's locations are collected at a large scale by a wide range of entities (e.g., \textit{Uber} and \textit{Google Maps}), typically through mobile technologies.  
Such data is extremely private, for numerous personal, social, and financial reasons.  
However, being able to analyze and model location patterns is highly valuable to other businesses and researchers (and society as a whole) to enable a vast range of location-based applications, from tracking disease spread to reducing traffic congestion.
The exponential growth in popularity of (open) data science has seen an ever-growing demand for the publication of a variety of location datasets (e.g., geotagged Tweets, taxi journey origins and destinations, social media check-ins).  
However, the risks concerning the violation of individuals' privacy present a major impediment to the free sharing of such data. 
Instead, the raw data has to be significantly sanitized before it can be published. 
This can involve aggregation into predefined regions, location perturbation, or truncation of longitude-latitude data.  
In this setting, the sanitization operation is controlled and performed by the data owner, whose primary concern is to minimize the privacy risk to the data subjects and their consequent liability. 
In many cases, this considerably limits the utility of the published data.

In contrast to crude sanitization, releasing a synthetic dataset in the same format as the original data can give more flexibility in how clients can use the published data. 
In many practical scenarios, the recipient of a dataset will want to use their in-house data analytics tools without any restrictions from the data provider on the way in which the data can be used, or the type of queries that can be asked.  
In this paper, we develop approaches for generating realistic synthetic data from real location data, while also satisfying the strict requirements of differential privacy (DP).
The aim is to maximize the similarity between the original and synthetic datasets, whilst protecting the existence and location of any individual.

Existing approaches to synthetic location data generation (surveyed in Section~\ref{s:related-work}) are unsatisfying for a number of reasons. 
They tend to adopt relatively simplistic ways to represent the data, such as fixed grids, and only materialize the population of cells within such grids.
They make crude uniformity assumptions within such basic regions that do not capture realistic location distribution patterns. 
They also tend to be oblivious of real-world conditions, such as straits of water or uninhabitable terrain, leading to nonsensical outputs that `locate' people in the middle of the ocean.  
In this paper, we propose novel solutions that overcome these limitations. 

Our first approach for synthetic data generation (SDG) targets the first of these weaknesses, by considering a richer set of ways with which to model the input location data. 
We introduce a differentially private partitioning-based framework in which we restrict SDG to be within small private regions. 
We introduce grid- and clustering-based methods, where we generate synthetic points within private regions using a novel adaptation of kernel density estimation that is specifically suited to our setting of multiple point generation and maintains privacy.
In all steps, privacy is provided by using DP mechanisms to add noise to counts, and it is maintained through the post-processing properties of DP.

In our second approach, we incorporate `common knowledge' about the world within the data generation process.
Traditionally, DP approaches make very restrictive assumptions regarding what outside knowledge is known beyond the data itself (e.g., provenance, structure, or hierarchy).  
However, it is common for a dataset to be strongly restricted or influenced by an underlying structure -- the nature or behavior of which is known to all.  
For example, location data is heavily influenced by the underlying road network, which is public knowledge.  
Our work is the first, to our knowledge, to exploit this underlying structure in order to generate differentially private synthetic location data.  
We first match the data to the given features (e.g., road segments) and materialize summary statistics using DP mechanisms.
From this, we generate synthetic points along each segment using privacy-preserving micro-histograms to maintain the underlying distribution. 

We perform an extensive set of experiments using real datasets with varying degrees of underlying structure.  
Our solutions perform significantly better than alternative approaches (up to 28x more accurate, and 3.7x faster).
The proposed partitioning-based approach is preferred when the data is less well-aligned with the underlying network, or when network data is unavailable. 
The proposed network-based approach is extremely effective, especially when the location data is well-aligned with the underlying road network. It is also up to 37x faster than partitioning-based approaches. 

Our methods further improve the real-life accuracy and utility of the generated data by incorporating public knowledge, such as streets, coastlines, and rivers.
We also evaluate the practical utility of the synthetic data in answering range, hotspot, and facility location queries.
The experimental results show that the synthetic data produces high quality results for these queries, thus highlighting both the strength of our approaches and the potential for widespread, real-world deployment of DP. Visualization of the real and synthetic data also improves explainability and trust in DP results.

A summary of our main contributions are:
\begin{itemize}
    \item a novel methodology and two robust methods for generating private synthetic location data with excellent performance in a range of location analytics tasks; 
    \item a new approach of incorporating public graph data (e.g., the road network) to enhance utility of private synthetic data;
    \item a novel mechanism for differentially private kernel density estimation that is designed for multiple point sampling for synthetic data generation; and
    \item an extensive evaluation of privacy-preserving data generation yielding several practical insights.
\end{itemize}

The rest of the paper is organized as follows. 
After reviewing the literature  (Section \ref{s:related-work}), Section \ref{s:problem-setting} introduces the problem, discusses its privacy and utility trade-off, and gives a brief overview of DP and its properties.
We explain our synthetic data generation solutions in Sections \ref{s:partitioning-based-approach} and \ref{s:road-network}, and evaluate them in Section \ref{s:expts}.  
In Section \ref{s:expts}, we also use the generated data to answer various location analytics queries.
We conclude our work with Section \ref{s:conc}.
\vspace{-0.2cm}
\section{Related Work}
\label{s:related-work}
Since DP has become the state-of-the-art privacy model, it has been applied to many domains, including medical, financial, and social network data.  
Using DP for spatial data is a continued area of focus given the significance and sensitivity of location data.  
For example, previous work has developed differentially private spatial decompositions~\cite{Cormode2012}, released spatial histograms~\cite{Ghane2018}, and protected temporally correlated location data~\cite{Xiao2015}.

There is an increasingly large body of work on private trajectory publication \cite[e.g.,][]{Gursoy2018} and synthesis \cite[e.g.,][]{He2015, Gursoy2020}.  Although these appear to be complex variants of the location privacy problem, the solutions therein all produce outputs that correspond to arbitrary grid cells (which is not concordant with the format of the original data), whereas we generate co-ordinate data (i.e., the same form as the input data).  While one could extend these solutions to generate individual points (e.g., by using uniform sampling), we show in our work that achieving high-quality results by synthesizing exact locations (while preserving the underlying characteristics of the real data) is a significant challenge.
Furthermore, almost all existing works fail to fully utilize publicly-known information to boost utility at no cost to privacy.  Although the work of \citet{Naghizade2020} is `context-aware', it lacks privacy guarantees, and there remains a high risk of reidentification.  Other context-aware work \cite[e.g.,][]{Acharya2019, Chen2016} uses the local setting of DP, as well as relaxed privacy definitions, which makes them incompatible with our objectives.

Notwithstanding the above differences, the problem we study is a core issue of spatial data publication with many important applications, such as advertising and better provision of public services. Our methodology addresses several practical challenges for real-life use of DP and private location data generation that are not considered in (or the focus of) previous works. 
Our work uniquely combines all of the following:
\begin{enumerate*}[label={\alph*)}]
    \item satisfying the strict requirements of DP under all circumstances;
    \item generating synthetic datasets in the same format as the input datasets; 
    \item contextualizing in the real world by incorporating real-world knowledge (e.g., road networks); and
    \item evaluating the methods with popular location analytics tasks.
\end{enumerate*}

\section{Problem Setting}
\label{s:problem-setting}
Given a dataset containing the real locations of individuals, we aim to generate synthetic spatial point data that satisfies $\epsilon$-DP, and preserves as much as of the underlying distribution of the real data as possible.  Specifically, our objective is to \textit{protect the existence and location of each individual in the dataset} by using differential privacy.  
We use $p$ and $s$ to denote real and synthetic locations (in co-ordinate form), and $\mathcal{P}$ and $\mathcal{S}$ to denote the sets of real and synthetic locations, respectively.
In this section, we outline how we seek to balance privacy and utility.  
We also briefly outline the setting of our problem with respect to adversaries and assumed knowledge.

\subsection{Privacy}
\label{ss:privacy}
Even when a strong social motivation for data sharing or release exists (e.g., in contact tracing to help track disease spread), there remains a need for strong privacy protections.
The absence of a sufficiently strong privacy model can result in deanonymization \cite{Sweeney1997} or inference attacks \cite{Krumm2007}.
We use DP as it provides a strong level of protection, through a guarantee of plausible deniability, to all members of a dataset.  

\begin{definition}[$\epsilon$-differential privacy \cite{Dwork2006, Dwork2006a}]
A randomized mechanism $\mathcal{A}$ is $\epsilon$-differentially private if, for any two datasets $D$ and $D'$ differing by one element, and for all $y \in \operatorname{Range}(\mathcal{A})$, we have:
\begin{equation}
   \frac{\Pr[\mathcal{A}(D) = y]}{\Pr[\mathcal{A}(D') = y]} \leq e^\epsilon
   \label{eq:dp}
\end{equation}
\end{definition}

In other words, a mechanism that satisfies $\epsilon$-DP should return approximately similar results, even if a tuple, $t$, is added or removed from a dataset (i.e., $D' = D \pm t$).  
The Laplace mechanism is used to release the values of numeric functions of data \cite{Dwork2006a}.  
For a function $f$ acting on $D$, it adds random noise to the value of $f(D)$ such that:
\begin{equation}
    \mathcal{A}_f = f(D) + \operatorname{Lap}\left(\frac{\Delta_f}{\epsilon}\right) 
\end{equation}
where, $\operatorname{Lap}(\cdot)$ denotes the Laplace distribution, and the scale of the noise is set by the sensitivity of $f$, $\Delta_f = \max_{D, D'}\; |f(D) - f(D')|$.

\eat{
The exponential mechanism \cite{McSherry2007} is an alternative method for releasing differentially private output. For any dataset $D$ and output $y \in \mathcal{Y}$, the result of mechanism $\mathcal{A}$ is $\epsilon$-differentially private if one randomly selects $y$ such that:
\begin{equation}
    \Pr[\mathcal{A}(D) = y] = \frac{\exp\left(\epsilon q(D,y) /2\Delta_q\right)}{\sum_{y_i\in \mathcal{Y}}\exp\left(\epsilon q(D,y_i)/2\Delta_q\right)}
    \label{eq:exp-mech}
\end{equation}
where, $q(D,y)$ is a quality function, and $\Delta_q$ is the sensitivity of the quality function (defined as for $\Delta_f$).
}

The privacy properties of multiple mechanisms can be analyzed via a composition theorem~\cite{Dwork2014}.
Multiple mechanisms $\mathcal{A}_i$, each with a privacy parameter $\epsilon_i$, can be combined to form one $\epsilon$-differentially private mechanism with $\epsilon = \sum_i\epsilon_i$.  
Thus, we refer to $\epsilon$ as the privacy budget for a specific task (i.e., synthetic location data generation), and apportion it into pieces.  
In our work, we add noise in at most three places and divide our privacy budget across these steps, where each step has a privacy budget of $\epsilon_i$.  That is, $\epsilon = \epsilon_1 + \epsilon_2 + \epsilon_3$. 

Another property of DP is its robustness to post-processing \cite{Dwork2014}.  That is, we can transform the output from a DP mechanism without further privacy loss, unless we use extra knowledge about the input.
When we use the Laplace mechanism for a count query, post-processing permits rounding all values to the nearest integer, and all negative values to zero, with no adverse privacy implications.  

\subsection{Utility}
\label{ss:utility}
Our aim is to generate synthetic data that maximizes utility, while meeting the above privacy guarantees. 
We initially assess this through two measures: normalized cell error (NCE) and mean edge distance difference (MEDD).

For NCE, we divide the region into $L$ cells (giving the set $\mathcal{L}$), and obtain $c_{l}^{\real}$ and $c_{l}^{\synth}$ -- the number of points in each cell for the real and synthetic datasets, respectively. 
NCE is then defined as:
\begin{equation}
    NCE = \frac{1}{|\mathcal{P}|}\sum_{l\in\mathcal{L}} |c_{l}^{\real} - c_{l}^{\synth}|
    \label{eq:mse}
\end{equation}

While NCE quantifies the error between just the synthetic and real datasets, MEDD quantifies the error between the two datasets with respect to a graph -- here, the road network.  We use MEDD to quantify the preservation of network alignment of the synthetic points.
We define $d(p, e_p)$ to be the shortest distance from a point $p$ to its nearest edge $e_p$ (explained more in Section \ref{s:road-network}). MEDD is hence defined as:
\begin{equation}
    MEDD = \left| \frac{1}{|\mathcal{P}|}\sum_{p\in\mathcal{P}} d(p, e_p) - \frac{1}{|\mathcal{S}|}\sum_{s\in\mathcal{S}} d(s, e_s) \right|
\end{equation}
As we seek to establish practical data sharing mechanisms, we also assess utility through a range of location analytics tasks like range, hotspot, and facility location queries.  These utility measures are described more in Section~\ref{s:expts}.

\subsection{Adversaries and Assumed Knowledge}
\label{ss:adversaries}
We assume that the aim of an adversary is to \textit{identify the true location of a certain individual}. 
As our proposed methods make use of external knowledge (e.g., the road network), which is \textit{public knowledge}, we assume it can also be utilized by any adversary.
Given this aim, there are two primary adversary targets: membership inference and location identification.  
To provide protection in both regards, we use differential privacy -- a widely-used, `road-tested' technique with strong, demonstrable privacy guarantees.  
Through its definition (see Definition 1), each individual has a degree of plausible deniability with respect to their inclusion in the synthetic dataset (governed by a probabilistic bound; see Equation \ref{eq:dp}).  
This assures us that the output $\mathcal{S}$ does not provide the adversary with an advantage in determining the true location of an individual in the input. 
Adopting \textit{synthesis} of location data (as opposed to \textit{publication}) further weakens the relationship between real and synthetic points.

As we treat each point independently, each point has its own (composable) DP guarantee.  As such, our methods can be applied to trajectory data without adverse downstream consequences.  That is, it would not be possible to link individual points in the synthetic data and re-identify a real trajectory.

\section{Partitioning-Based Data Generation}
\label{s:partitioning-based-approach}

This section details our two-stage partitioning-based approach.  
We first restrict data generation to be within small regions, and then generate a noisy number of points, while preserving a distributional measure of the real data. 
We propose a private version of kernel density estimation (KDE) to obtain representative probability distributions of point data.
For the kernel function to be well-defined, it requires access to points in the database, which makes satisfying DP requirements difficult while maintaining high utility. 
Privatizing KDE is further complicated by our need to repeatedly sample from the private KDE to generate multiple synthetic points, a process that would potentially lead to high levels of privacy leakage ordinarily.  
Hence, we develop a kernel density estimate that satisfies $\epsilon$-DP, achieves high utility, and is robust to multiple sampling.

\subsection{Private Data Partitioning}
\label{ss:partition}

Before introducing our solution for generating data, we outline how we partition our space by using differentially private grid- and clustering-based approaches from the literature.

\subsubsection{Grid-Based Partitioning}
\label{sss:grid}
A simple method to privately partition data is to use a uniform grid (UGrid) that is independent of the data, thus maintaining privacy. Choosing the correct granularity, however, is important as too coarse or too fine a grid can lead to poor results.  Consequently, to determine the dimensions of the grid, we utilize a guideline proposed in~\citet{Qardaji2013}.  For an $m \times m$ uniform grid, we set the number of cells in each direction to be:
\begin{equation}
    m = \left\lceil\sqrt{\frac{N\epsilon}{10}}\right\rceil
    \label{eq:min-grid-size}
\end{equation}
where, $N$ is the number of points in the real dataset, $\mathcal{P}$, and $\epsilon_1$ is the privacy budget assigned to this task.  This ensures that the average number of points per cell is suitably larger than the noise magnitude, and it follows the composition property of DP introduced in Section \ref{ss:privacy}. 
Consequently, the total number of cells, or regions, into which the data is partitioned is $K = m^2 \approx \frac{N\epsilon_1}{10}$.  
We add noise to the number of points $n_i$ in each region $R_i$ using the Laplace mechanism to obtain: $n'_i = n_i + \operatorname{Lap}(\frac{1}{\epsilon_1})$.  

In many situations (e.g., non-uniform distribution of points), a uniform grid would be unsuitable as it would likely fail to capture the distribution accurately and/or add noise to the dataset in a biased manner.  Therefore, we also implement an adaptive grid (AGrid) method (from \citet{Qardaji2013}) whereby denser regions have more grid cells, and sparser regions have fewer cells.   We follow their recommendation by first dividing the data region into an $m_1 \times m_1$ uniform grid where:
\begin{equation} 
\label{eq:agrid1}
     m_1 = \max \left(10, \frac{1}{4}\left\lceil\sqrt{\frac{N \epsilon_1}{10}}\right\rceil\right)
\end{equation}
We add Laplace noise, controlled by $\epsilon_1$, to the count in each cell and then divide each cell $i$ into an $m^i_2 \times m^i_2$ grid where:
\begin{equation} 
    \label{eq:agrid2}
     m^i_2 = \left\lceil\sqrt{\frac{n'_i\epsilon_2}{5}}\right\rceil
\end{equation}
We conclude the partitioning phase by adding Laplace noise, controlled by $\epsilon_2$, to the count in each of the new smaller cells.  

\subsubsection{Cluster-Based Partitioning}
\label{sss:cluster}
We also implement a private clustering-based approach to generate regions.
\eat{Most early DP approaches in the literature require substantial effort to handle issues concerning the initial selection of centroids, number of necessary iterations, and the addition of necessary noise to the data \cite{Blum2005, Zhang2013, Balcan2017, Biswas2019, McSherry2009, Mohan2012}.}
We adapt the expanded uniform grid $K$-means (EUG$K$M) method \cite{Su2016, Su2017}, which has been shown to perform well while satisfying $\epsilon$-DP.  

In short, EUG$K$M consists of two steps: initial cluster centroid generation and $K$-means-style clustering.   
To generate the locations of an initial set of $K$ centroids, EUG$K$M uses the concept of sphere packing to randomly generate points within the bounds of the dataset that ensures that all centroids are evenly (but not necessarily equally) spaced across the data space. 
The main advantage of this method is that it can be done without access to individual data records, thus maintaining privacy.
A uniform grid is then generated using Equation \ref{eq:min-grid-size}, and $\epsilon_1$ is used to control the grid size.  
Data points are assigned to a grid cell, the total number for each cell is calculated, and Laplace noise of $\operatorname{Lap}(\frac{1}{\epsilon_1})$ is added to the count in each cell.  
Grid cells are then `allocated' to their nearest centroid and a weighted $K$-means style procedure for optimization is initiated, where the cell-centroid distances are weighted by the (noisy) number of points in each cell.
We use these centroid locations to generate $K$ Voronoi regions to which each real data point is assigned.  
For each cluster region, we obtain the number of points and, as we have interacted with the real data again, we need to add noise to each Voronoi region's count.
Hence, our final step is to add Laplace noise to get a noisy count: $n'_i = n_i + \operatorname{Lap}(\frac{1}{\epsilon_2})$.
Once again, this is in accordance with DP's composition property  (Section \ref{ss:privacy}).

In summary, the main difference between the two partitioning methods is that clustering is (in theory) more sensitive to non-uniform point distributions (i.e., using Voronoi regions allows small clusters to form easily in dense regions).  We examine this empirically in our experiments.

\subsection{Private Data Generation}
\label{ss:data-gen-approaches}
Generating synthetic data from a domain without imposing any constraints can be done in many ways.  
For example, sampling from a uniform distribution over the entire domain will maximize the entropy. 
However, we aim to generate synthetic data that preserves some underlying characteristics or properties of the real data.

Our task is made more difficult as we try to match more complex features of the data while imposing the strict requirements of $\epsilon$-DP.

In this section, we introduce differentially private SDG methods for use in conjunction with any partitioning method. 
Note that, when generating synthetic points with any method, we can ensure that points are not generated in regions that are unlikely to contain points, such as seas and rivers. 
We do this by specifying `out-of-bounds' regions from which we filter any synthetic data points that lay within these regions.  
More explanation of this process is given in Section~\ref{ss:outline}.

\subsubsection{Uniform Distribution}
\label{sss:uniform}
As private partitioning already approximately captures an overall distribution of the points, a simple method for synthetic point generation is to sample at random from a uniform distribution. 
As uniform random sampling is data independent, no further noise is needed at this stage to preserve privacy (i.e., $\epsilon_3 = 0$).  We further reduce the size of the region by dividing each region into triangles where each triangle consists of the region's centroid and two adjacent vertices of the region. We generate points randomly within each triangle in proportion to each triangle's area, using the triangle point picking method~\cite{Weisstein2020}. 

\subsubsection{Weighted Uniform Distribution}
\label{sss:kdd}
A more nuanced approach is to use information from neighboring regions to define the point distribution.  
The \textit{weighted uniform distribution} (WUD) approach sub-divides each region and distributes points uniformly across each sub-region. 
The number of points in each sub-region is influenced by characteristics of the sub-region and neighboring region~\cite{Yilmaz2017}.  

We split each region $R_i$ into $J$ sub-regions. 
The number of points $n'_{i,j}$ in sub-region $R_{i,j}$ is
based on its area and the noisy number of points in the neighboring region.  It is defined as:   
\begin{equation} 
    n'_{i,j} = n'_i\left(\omega \frac{A_{i,j}}{A_i} + (1-\omega) \frac{x'_{i,j}}{x'_i}\right)
\end{equation}
where, $A_{i,j}$ and $A_i$ are the areas of $R_{i,j}$ and $R_i$, respectively; $x'_{i,j}$ and $x'_i$ are the noisy number of points in the neighboring region(s) to $R_{i,j}$ and $R_i$, respectively; and $0 \le \omega \le 1$ is a weighting factor.  
By definition, $x'_i = \sum_j x'_{i,j}$.  
We set $\omega=0.5$ to give equal weight between the areas and populations of (sub-)regions.  
Once the number of points in each sub-region is determined, we generate points using the triangle method (Section \ref{sss:uniform}).
As the boundary regions are private (due to the partitioning method) and we only ever use the noisy number of points in any region, the post-processing property of DP negates further noise addition. Hence, $\epsilon_3 = 0$ here.

\subsubsection{Kernel Density Estimation}
\label{sss:kde}
Kernel density estimation is a statistical approach to estimate the density function of a distribution.  
Using KDE as a basis for synthetic data generation can better preserve the underlying characteristics of the original data.

The kernel density estimator, ${\hat{f}}(\mathbf{x})$, is defined as: 
\begin{equation}
    \hat{f}(\mathbf{x})={\frac{1}{N}}\sum_{j=1}^{N}\mathcal{\phi}(\mathbf{x}-\mathbf{x}_{j})
    \label{eq:kde}
\end{equation}
where, $\mathbf{x}$ is a two-dimensional vector consisting of $x$- and $y$-co-ordinates, $N$ is the number of points in the dataset (that is the basis for the kernel), and $\phi$ is the kernel function.  

\para{Kernel Density Estimator Construction}
While there have been numerous attempts to privatize KDE \cite{Hall2013, Alda2017, Huai2019}, these methods are not well-suited to our setting (i.e., sampling multiple times from a private KDE). Prior efforts adopt relaxed privacy definitions, such as $(\epsilon,\delta)$-DP \cite{Hall2013}, or perform post-hoc testing of KDE samples for privacy \cite{Huai2019}.  
\citet{Alda2017} use the Gaussian kernel, which results in oversmoothing in our setting, leading to poor quality synthetic data.

We instead use a two-dimensional Laplace kernel, owing to the widespread use of its one-dimensional counterpart in other DP work.  
Specifically, we use the polar Laplace distribution, which has the probability density function:
\begin{equation}
    \phi(\mathbf{x}-\mathbf{x}_{j}) \equiv \phi(r, \theta) = \frac{\exp(-r/h)}{2\pi h}
    \label{eq:kernel}
\end{equation}
where $r = \|\mathbf{x}-\mathbf{x}_{j}\|$, $\theta$ is the angle between $\mathbf{x}$ and $\mathbf{x}_{j}$, and $h$ is a normalization (or smoothing) factor.
To ensure we obtain a differentially private kernel for region $R_i$, it is necessary to tune the kernel function in each region $R_i$ such that the probability ratio between the two most distal points in $R_i$ is no more than $e^\epsilon$, as required by Definition 1. 
Hence, we set the smoothing parameter for $R_i$ to be:
\begin{equation}
    h_i = \frac{\|R_i\|}{\epsilon*}
    \label{eq:h}
\end{equation}
where $\|R_i\|$ is the maximum distance between any two locations (not necessarily in $\mathcal{P})$ in $R_i$. 
Consequently, proving that this kernel function satisfies DP can be easily done by examining the probability ratio between $\phi(0, \theta)$ and $\phi(\|R_i\|, \theta)$.

\para{Synthetic Data Generation}
We now outline how to generate a synthetic point $s$.
To do so, we utilize a convenient property of kernel density estimation: sampling from the full KDE is equivalent to first sampling one of the $n$ points $\mathbf{x}_j$, then sampling from the kernel around $\mathbf{x}_j$.
From Equation \ref{eq:kernel}, we see that $r$ and $\theta$ can be sampled independently -- that is, $\phi(r, \theta) = \phi(r)\phi(\theta)$.  
To this end, we first sample from 
$\phi(r) = h^{-1}\exp(-r/h)$
, and then sample from $\phi(\theta) = 1/2\pi$ (equivalent to sampling randomly from the uniform distribution with bounds $(0, 2\pi])$. 
Once we obtain values for $r$ and $\theta$, we convert to Cartesian co-ordinates and add this displacement to the sampled real point to give $s$ (i.e., $x_s = x_p + r\cos{\theta}, y_s = y_p + r\sin{\theta}$).

There is a risk that real points are sampled many times, which would lead to privacy leakage that could reveal the true location of an individual.  
To avoid this, we modify the sample procedure slightly. 
We set $\epsilon^* = \epsilon_3 /\lambda$, which allows each real point to be sampled at most $\lambda$ times (using sequential composition), meaning we achieve our target level of privacy protection.
If we reach this limit, or if $n_i = 0$ and $n'_i > 0$, we simply generate a point uniformly at random, which has no negative privacy consequences.
We set $\lambda = 2$ as $n'_i \leq 2n_i$ in most cases.
We repeat this sampling process until $n'_i$ points are generated in each region $R_i$.
Finally, as a sample generated this way has the same distribution as the KDE and the KDE satisfies DP, it follows that the synthetic data satisfies DP.
\section{Road Network- and Geography-Aware Data Generation}
\label{s:road-network}

The methods presented thus far follow the common assumption that there is limited knowledge of the underlying geography.
In many cases, however, more significant information is available both to the data owner and to the public. 
For example, for a dataset of vehicle trajectories, it is reasonable to assume that all points in the dataset will correspond to points on (or very close to) segments of a city's road network.  
Therefore, when generating points, one should ensure that all synthetic points are similarly aligned to road segments.  
We can also use outside knowledge to infer where individuals may be unlikely to be located (e.g., in seas, rivers, military bases). 
Importantly, enforcing these constraints does not use any information not already in the public domain, and can therefore be done without using any of the privacy budget. 
For example, the location of roads and boundaries of seas are available (often to a high level of detail) through a range of mapping platforms and government open data repositories.

\begin{figure}[t]
    \centering
    \includegraphics[height = 3cm]{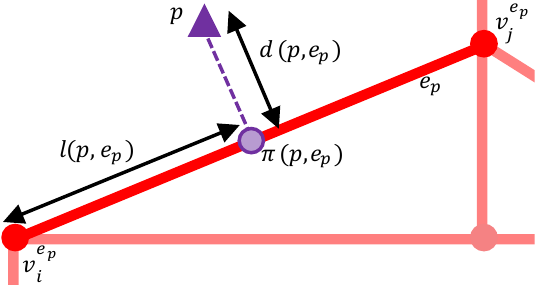}
    \caption{Diagram showing \textit{p}, \textit{e\textsubscript{p}}, \textit{d(p, e\textsubscript{p})}, and \textit{l(p, e\textsubscript{p})}}
    \label{fig:network-variable-pic}
\end{figure}

\para{Notation}
Consider the graph $G(E,V)$ that represents the road network.
$E$ and $V$ represent the road segments and road intersections, respectively.  
For each individual location $p \in \mathcal{P}$, there exists an edge $e_p \in E$ that is the closest edge (distance-wise) to $p$.
Two distance functions help us map $p$ onto $e_p$
(see Figure \ref{fig:network-variable-pic}). 
The first, $d(p, e_p)$ gives the perpendicular distance from $p$ to $e_p$.  
The projection of $p$ onto $e_p$  is denoted by $\pi(p, e_p)$. 
The second function, $l(p, e_p)$, gives the distance along $e_p$ between $v_i^{e_p}$ and $\pi(p, e_p)$.  

\para{Noise Addition}
If the real data points are not perfectly aligned with the assumed road network, it is necessary to map-match them to edges in the graph (i.e., obtain $e_p$ for all $p \in \mathcal{P}$).  
We count the number of points for which that edge is the nearest, and denote it as $n_e$.  
We now use this count to determine the noisy number of points that will be generated along each edge by first adding Laplace noise to $n_e$. 
The privacy budget is represented as $\epsilon_1$, using the composition property of DP (see Section \ref{ss:privacy}).  A simple approach would be to use these values as the noisy counts.  However, this would result in a large amount of additional noise throughout the dataset, especially when a large proportion of edges have low/zero counts.
Therefore, we reduce the influence of the noise by denoting this `intermediate' count as $n^*_e$, and performing a post-processing step to obtain $n'_e = \frac{N \times n^*_e}{N^*}$, where $N^* = \sum_e n^*_e$, the sum of intermediate noisy counts for all edges.  Furthermore, we set $n'_e = 0$ for all edges where $n'_e \leq \theta$, where $\theta$ is a threshold value.  Imposing this threshold also reduces the impact of the added Laplace noise.  
DP is still satisfied as these are post-processing operations.

\para{Determining the threshold value}
The value of $\theta$ can impact the quality of the synthetic data and may vary dynamically with $\epsilon_1$ (as the magnitude of added noise depends on $\epsilon_1$).  
The optimal value for $\theta$ 
will balance
the number of points added to edges where $n_e = 0$, and the number of points `lost' for edges where $n'_e \leq \theta$ and $n_e \neq 0$.  
However, trying to find this equilibrium directly requires knowing the true number of points on each edge, which would violate DP.  
To obtain a good approximation for $\theta$, we use the inverse cumulative distribution function of the Laplace distribution, defined as:
\begin{equation}
\label{eq:quantile}
Q = 
\begin{cases} 
    \mu + \frac{\text{ln}(2F)}{\epsilon_1} & \mbox{if } F \leq 0.5 \\
    \mu + \frac{-\text{ln}(2-2F)}{\epsilon_1} & \mbox{if } F \geq 0.5
\end{cases}
\end{equation}
where, $Q$ is the quantile of the Laplace distribution, $\mu$ is the mean of the distribution (i.e., $n_e$), and $F$ is the value of the cumulative distribution function.  
The intuition is that setting $\theta = Q_{95}$ seeks to remove approximately 95\% of the added noise, for example.  
When $n_e = 0$, then $\mu = n_e = 0$ and so $Q = 0$ when $F \leq 0.5$ (disbarring negative counts), so we only need the second term. 
Furthermore, when $\epsilon_1$ is very small, the above term can be very large, which also causes adverse distortion to the dataset.  
Thus, we impose an upper limit on the value $\theta$ can take, which we set to be 10.  
Hence, $\theta$ is defined as:
\begin{equation}
\label{eq:theta}
    \theta = \min \left(\frac{-\text{ln}(2 - 2F)}{\epsilon_1}, 10\right)
\end{equation}
Experimentally, we find $F=0.9$ (i.e., removing about 90\% of the noise) to be satisfactory, so this is our default choice. 

\begin{figure}[t]
    \centering
    \includegraphics[height = 3cm]{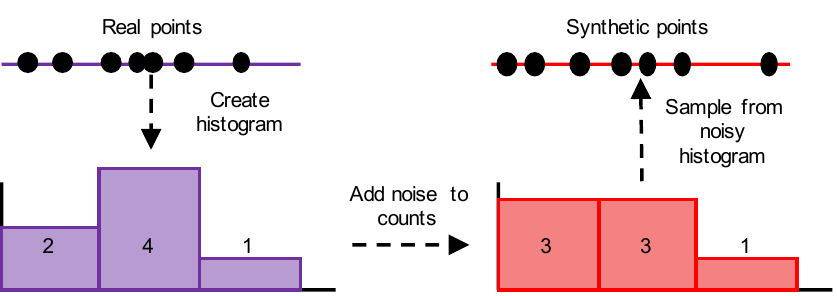}
    \caption{Example for generating the values for \textit{l(s, e\textsubscript{s})}}
    \label{fig:along-edge-example}
\end{figure}

\para{Synthetic Data Generation}
To generate a synthetic point $s$ along an edge, we must fix (i) the distance along $e$ that $s$ is, (ii) the perpendicular distance from $e$ that $s$ is, and (iii) the `side' of the edge that $s$ is in relation to $e$.
For (i), we could assign a distance at random from a uniform distribution.  
However, for very long roads, this could result in synthetic points being far from the real point locations, which would possibly reduce the synthetic data's utility.  
Instead, we summarize each edge with a micro-histogram.  
For each edge, we create a histogram (with $\alpha$ bins) using the values of $l(\cdot, e_p)$ and, to preserve privacy, we add noise ($=\operatorname{Lap}(\frac{1}{\epsilon_2})$) to the count of each bin.
We sample from this noisy histogram to determine the bin in which $s$ lies, and the exact value for $l(s, e_s)$ is determined by sampling from a uniform distribution with bounds corresponding to the bounds of the histogram bin.  
We sample from the histogram $n'_e$ times to generate the necessary values for $l(s, e_s)$ -- note that $e_s \equiv e_p$.  
A pictorial example of this process is shown in Figure \ref{fig:along-edge-example}. 

For (ii), we use the same approach to determine the values of $d(s, e_s)$, with $\epsilon_3$ as the privacy budget when adding noise to the histogram.
When the values for $d(s,e_s)$ and $l(s, e_s)$ are set, there are two possible locations for $s$.  
For (iii), we select between these two locations with equal probability to determine the final location of $s$.  
When $n_e = 0$, we define the range of histogram values such that $d(s, e_s)$ takes a value in the range (0, 10) meters and $l(s, e_s)$ takes a value in the range (0, $|e_s|$), where $|e_s|$ is the length of $e_s$.
This process is applied to all edges in $E$ where $n'_e >$ 0 until the entire synthetic dataset, $\mathcal{S}$, is created.

\begin{table*}[t]
  \caption{Dataset Information}
  \centering
  \label{tab:datasets}
  \begin{tabular}{cccc|cccc|c}
    \toprule
    \multirow{2}{*}{\textbf{City}} & \multirow{2}{*}{\textbf{Data Type}} & \textbf{Number of} & \textbf{Number of} & \multicolumn{4}{c|}{\textbf{Boundaries}} & \multirow{2}{*}{\textbf{Ref.}} \\
    & & \textbf{Points} & \textbf{Edges} & \textbf{North} & \textbf{South} & \textbf{West} & \textbf{East} & \\
    \midrule
    New York, USA  & 311 Calls & 163,220 & 8,161 & \multicolumn{4}{c|}{Manhattan Island} & \cite{NYC2020} \\
    Beijing, China      & Taxi Trajectories & 158,260 & 7,913 & 39.954 & 39.862 & 116.330 & 116.450 & \cite{Yuan2010, Yuan2011} \\
    Porto, Portugal     & Taxi Trajectories & 79,360 & 3,968 & 41.168 & 41.123 & -8.635 & -8.576 & \cite{Porto2015} \\
  \bottomrule
\end{tabular}
\end{table*}

\para{Histogram bin choice}
\label{sss:histogram}
We now discuss how to choose $\alpha$, which affects the downstream utility of the synthetic data. 
We aim to balance the amount of overall noise added to an edge with the location accuracy along an edge.
For example, having a high number of bins will be beneficial for 
describing locations accurately, but will involve high noise addition, which will negatively affect the accuracy during the histogram sampling stage.  The converse is true for low $\alpha$.
\begin{theorem}
The optimal value of $\alpha$ is $\mathcal{O}\left(\sqrt{\epsilon N}\right)$.
\end{theorem}
\begin{proof}
Consider a road segment with $N$ points that is divided into $\alpha$ histogram bins.  
Suppose we have a range count query that covers a proportion $q$ of the road (i.e. $q\alpha$ bins).  
The error in answering a range query has two components: privacy noise error and non-uniformity error.  
Knowing that the expected magnitude of the $\epsilon$-DP noise error per bin is $\sqrt{2}/\epsilon$, 
we know that the noise error will be proportional to $\frac{\sqrt{2}q\alpha}{\epsilon}$.  
There are, on average, $\frac{N}{\alpha}$ points in each bucket.  
When answering a query that partially intersects a bin, we do not know whether points in a bin will be included in the query response (owing to  non-uniformity of the data distribution), and this uncertainty is $n_j$, the number of points in the $j$\textsuperscript{th} bin. 
Summing over queries touching each of $\alpha$ bins, the total error is given by:
$\sum_{j=1}^{\alpha} j \frac{\sqrt{2}}{\epsilon} + n_j = \frac{\alpha(\alpha+1)\sqrt{2}}{2\epsilon} + N$. 
We minimize this error by equating the two terms, yielding 
$\alpha = \mathcal{O}\left(\sqrt{\epsilon N}\right)$.
\end{proof}

In this proof, $N$ corresponds to $n'_e$, the noisy count of the edge; and $\epsilon$ corresponds to either $\epsilon_2$ or $\epsilon_3$.
The value for $\epsilon$ can be chosen empirically and we find that setting $\alpha = \sqrt{N}$ gives effective results, as demonstrated in the experiments.

\begin{figure*}[t]
    \centering
    \begin{subfigure}[b]{0.35\columnwidth}
        \centering
        \includegraphics[width=\textwidth]{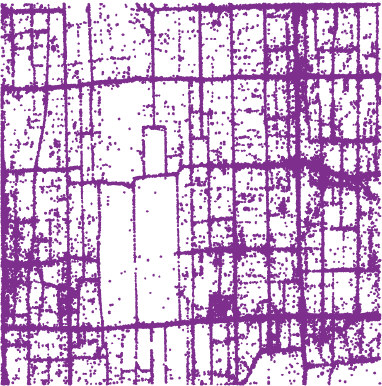}
        \caption{Real Points}
        \label{fig:real-points}
    \end{subfigure}
    \hfill
    \begin{subfigure}[b]{0.35\columnwidth}
        \centering
        \includegraphics[width=\textwidth]{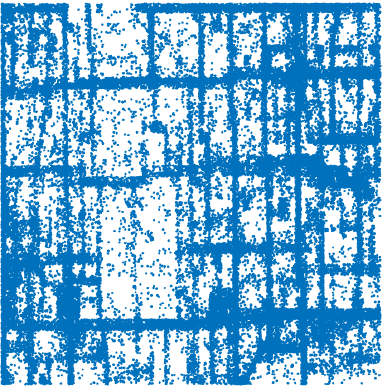}
        \caption{UGrid-KDE}
        \label{fig:grid-points}
    \end{subfigure}
    \hfill
    \begin{subfigure}[b]{0.35\columnwidth}
        \centering
        \includegraphics[width=\textwidth]{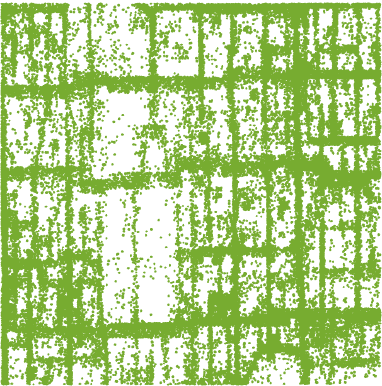}
        \caption{AGrid-KDE}
        \label{fig:agrid-points}
    \end{subfigure}
    \hfill
    \begin{subfigure}[b]{0.35\columnwidth}
        \centering
        \includegraphics[width=\textwidth]{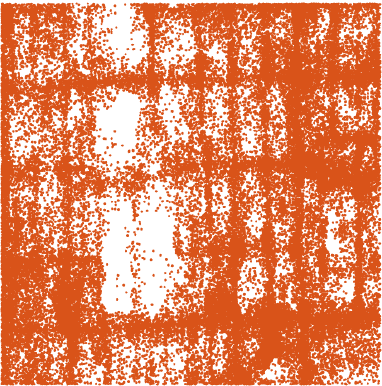}
        \caption{Clust-KDE}
        \label{fig:cluster-points}
    \end{subfigure}
    \hfill
    \begin{subfigure}[b]{0.35\columnwidth}
        \centering
        \includegraphics[width=\textwidth]{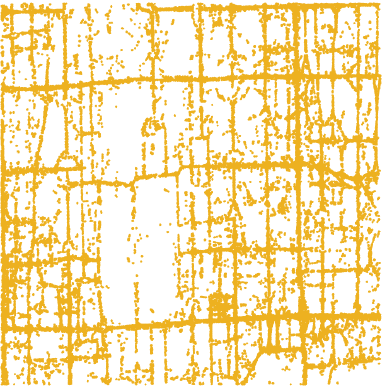}
        \caption{Road}
        \label{fig:road-points}
    \end{subfigure}
    \caption{Plots of real and synthetic data for data generation methods (Beijing)}
    \label{fig:visual-results}
\end{figure*}
\section{Experimental Evaluation}
\label{s:expts}

In this section, we assess the accuracy and efficiency of our methods using the utility measures from Section~\ref{ss:utility}. 
We also evaluate our synthetic datasets for a range of common location analytics tasks. 
We outline our experiments in Section \ref{ss:outline}, before comparing our synthetic data generation methods in Section \ref{ss:comparison-methods}.  We then consider our application-focused queries: range and hotspots queries in Section \ref{ss:range-hotspot}, and facility location queries in Section \ref{ss:fl}.  We finish the section with discussion and recommendations (Section \ref{s:discussion}).

\subsection{Experiment Outline}
\label{ss:outline}

\para{Datasets}
We generate synthetic data using real location data from three cities with different topographies and sizes, detailed in 
Table~\ref{tab:datasets}.

We extract only the longitude-latitude pairs of each record.  
Although taxi trajectory points are correlated, we consider each point to represent an independent individual in the dataset.  We ignore any temporal information connected to the experiment data.  
We extract coastline data from OpenStreetMap and use this to define `out-of-bounds' regions that represent major bodies of water, such as seas and rivers.  
We remove any points in the original data located in these out-of-bounds regions, and  ensure that no synthetic points are created in these regions.
The same technique can be used to add further geographical restrictions (e.g., forests, military bases) on the presence of real or synthetic individuals.

We extract the `driveable' road network data as a graph from OpenStreetMap, using the \texttt{osmnx} Python package \cite{Boeing2017}, with boundaries matching those detailed in Table \ref{tab:datasets}. As pre-processing steps, we map-match each point in the cleaned datasets to the corresponding road network, remove any edges that are within the out-of-bounds areas, and  calculate the values $d(p, e_p)$ and $l(p, e_p)$ for each point.  The final number of points and edges for each city is shown in Table \ref{tab:datasets}.  
To examine the real-world suitability of our methods, we do not correct the map-matched data to enforce alignment with the road network; we discuss this more in Section~\ref{sss:real-world-considerations}.

\para{Baselines}
As discussed in Section \ref{s:related-work}, most existing work only publishes count data for grid cells/clusters, as opposed to generating co-ordinate data.  
Such data can be generated in these partitions using simple uniform sampling (Section \ref{sss:uniform}) and so we use these extensions of existing methods as baselines.  We use the terms `UGrid-Uni', `AGrid-Uni', and `Clust-Uni' to refer to the extension of the uniform grid, adaptive grid, and clustering-based partitioning methods.

\para{Parameter Selection}
For each dataset, we set the number of data points, $N=20|E|$, where $|E|$ is the number of edges.
We do this so that the number of grid cells is approximately equal to the number of edges for the road network-based solution.  (We refer to this method simply as `Road'.)
This allows for a fairer comparison between the methods as the amount of added noise will be more comparable.  
However, for clustering-based methods, having $K \approx |E|$ would result in the regions exhibiting a grid-like structure, and so we set $K = 1{,}000$.  
By default, $\epsilon = 1$, but we evaluate the impact of varying and splitting the privacy budget in Section~\ref{sss:varying-parameters}.

\para{Utility Measures}
\label{sss:utility}
We use the two measures detailed in Section \ref{ss:utility}: normalized cell error (NCE) and mean edge distance difference (MEDD).
To calculate the NCE, we divide the entire region into a uniform grid where each individual grid cell has approximate real-life dimensions of 100m $\times$ 100m.

\begin{table*}[tb]
    \centering
    \caption{NCE, MEDD, and runtime values for default settings; baselines denoted by asterisks (*)}
    \label{tab:similarity-results}
    \begin{tabular}{cc|ccc|ccc|ccc}
        \toprule
        \multicolumn{2}{c|}{\textbf{Data Generation}} &
        \multicolumn{3}{c|}{\textbf{Beijing}} & \multicolumn{3}{c|}{\textbf{Porto}} & \multicolumn{3}{c}{\textbf{New York City}} \\
        \multicolumn{2}{c|}{\textbf{Method}} & 
        \textbf{NCE} & \textbf{MEDD} & \textbf{Time} & 
        \textbf{NCE} & \textbf{MEDD} & \textbf{Time} &
        \textbf{NCE} & \textbf{MEDD} & \textbf{Time} \\
        \midrule 
        \multirow{3}{*}{UGrid} & Uniform* & 0.360 & 10.64 & 64.25 & 0.165 & 6.46 & 62.58 & 0.374 & 15.36 & 91.44 \\ 
         & WUD & 0.332 & 8.59 & 295.97 & \textbf{0.152} & 5.36 & 131.62 & 0.366 & 15.04 & 233.39 \\ 
        & KDE & 0.297 & 8.62 & 39.89 & 0.160 & 5.22 & 99.63 & 0.309 & 12.86 & 749.28 \\ 
        \midrule 
        \multirow{3}{*}{AGrid} & Uniform* & 0.379 & 11.83 & 55.92 & 0.188 & 6.23 & 65.33 & 0.310 & 14.99 & 159.19 \\ 
        & WUD & 0.362 & 10.02 & 1336.85 & 0.180 & 5.20 & 399.73 & 0.307 & 14.63 & 1469.85 \\ 
        & KDE & \textbf{0.285} & 8.84 & 63.82 & 0.160 & 4.76 & 265.71 & 0.259 & 11.34 & 1876.09 \\ 
        \midrule 
        \multirow{3}{*}{Cluster} & Uniform* & 0.876 & 27.83 & 10.81 & 0.407 & 13.00 & 17.51 & 0.610 & 19.48 & \textbf{25.17} \\ 
        & WUD & 0.866 & 26.19 & 28.88 & 0.391 & 12.38 & 32.50 & 0.591 & 18.63 & 29.11 \\ 
        & KDE & 0.616 & 19.23 & \textbf{8.23} & 0.272 & 8.85 & 85.93 & 0.463 & 16.41 & 842.54 \\ 
        \midrule 
        \multicolumn{2}{c|}{Road} & 0.316 & \textbf{1.97} & 29.09 & 0.184 & \textbf{0.94} & \textbf{16.87} & \textbf{0.200} & \textbf{0.70} & 51.40 \\
        \bottomrule
    \end{tabular}
\end{table*}

\subsection{Comparison of Methods}
\label{ss:comparison-methods}

\subsubsection{Summary}
\label{sss:main-results}
Figure \ref{fig:visual-results} shows the visual similarity between the real and synthetic data.
Although all methods preserve the underlying structure to some degree, we see that utilizing the geographical constraints explicitly in the SDG stage produces synthetic data that has much stronger visual similarity to the real data than the partitioning-based methods.
Quantitatively, Table~\ref{tab:similarity-results} shows the NCE and MEDD values for the four SDG methods, as well as the three approaches for generating synthetic data points within defined regions (Section \ref{ss:data-gen-approaches}) and the runtimes for each. 

Adopting KDE for grid-based partitioning methods improves data quality, compared to extensions of existing methods.  
As KDE almost always outperforms WUD in accuracy terms, we adopt it as the default choice for data generation.  
AGrid performs similarly to UGrid, unless the city's network is more structured (e.g., New York) in which case it is markedly better.  
We note that it (generally) takes longer to run, which may make UGrid preferable.
For clustering-based partitioning, KDE offers notable improvements compared to other approaches, although it fails to match the grid-based approaches in accuracy terms.  
This is primarily because 
larger regions lead to flatter kernels due to the requirements of DP (i.e, $\|R_i\|$ is larger, meaning $h$ is larger).

Using Road offers even greater improvements, as we observe improvements of up to 28x over the baselines (vs. Clust-Uni, MEDD, New York).  Furthermore, Road is up to 3.9x faster than the baselines (vs. AGrid-Uni, Porto), and up to 37x faster than KDE approaches (vs. AGrid-KDE, New York).  This highlights its suitability for generating large city-scale synthetic datasets of high utility. 

In Porto and Beijing, where many points are not closely aligned with the road network and the road network is less ordered, grid-based approaches are generally superior in accuracy terms. 
In New York, however, the real data adheres more tightly to the road network, which means Road is much better at creating high quality synthetic data and it achieves better MEDD values.

\subsubsection{Varying Parameters}
\label{sss:varying-parameters}
We also examine the effects that varying the key parameters have on the quality of the data.  
Owing to space limitations, we omit some corresponding plots.

\begin{figure}[!t]
    \centering
    \begin{subfigure}[b]{\columnwidth}
        \centering
        \includegraphics[width=4.5cm]{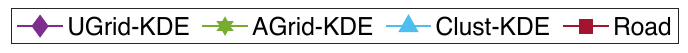}
        \label{fig:l4gend4}
    \end{subfigure}
    \\
    \begin{subfigure}[b]{0.49\columnwidth}
        \centering
        \includegraphics[height = 4.5cm]{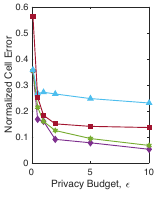}
        \caption{NCE}
        \label{fig:pb-mse}
    \end{subfigure}
    \hfill 
    \begin{subfigure}[b]{0.49\columnwidth}
        \centering
        \includegraphics[height = 4.5cm]{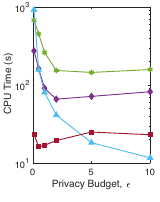}
        \caption{Time}
        \label{fig:pb-time}
    \end{subfigure}
    \caption{Variation in NCE and CPU time with $\bm{\epsilon}$ (Porto)}
    \label{fig:privacy-budget}
\end{figure}

\para{Privacy Budget}
Figure \ref{fig:privacy-budget} shows the effect of changing $\epsilon$ on the NCE and runtime (for Porto, although other cities exhibit similar profiles).  In terms of accuracy, all methods behave as expected: accuracy decreases as $\epsilon$ decreases, due to the increase in the amount of added noise.  
For low $\epsilon$, runtime is higher  for partitioning-based methods as it is more likely that generated points are `out-of-bounds' or outside the boundaries of $R_i$.  
Runtimes for grid-based methods increase as $\epsilon$ increases beyond 5 as the number of cells grows in proportion to $\epsilon$ (cf. Equations \ref{eq:min-grid-size}-\ref{eq:agrid2}).
The runtimes for Road are consistently low for all $\epsilon$, which further highlights its general suitability.

\para{Privacy Budget Distribution}
To examine the effect of varying the distribution of $\epsilon$, we consider the following apportionments.
First, note that, for all UGrid methods, $\epsilon_2 = 0$ as noise is only added once during the partitioning phase.
Likewise, for data generation methods that do not use KDE, recall that $\epsilon_3 = 0$.
Hence, for UGrid methods that do not use KDE, $\epsilon_1 = \epsilon$.
For UGrid methods with KDE-based data generation, we consider the following percentage splits between $\epsilon_1$ and $\epsilon_3$: 10--90; 20--80; 30--70; 40--60; 50--50\, and their reverses.  We find that, empirically, the best privacy budget split is to set $\epsilon_1 = 0.6\epsilon$ and $\epsilon_3 = 0.4\epsilon$.  
This is intuitive as it achieves approximate balance in noise addition between the partitioning and data generation phases.
For AGrid partitioning, we follow the guidance in \citet{Qardaji2013} and set $\epsilon_1 = \epsilon_2$.
For KDE-based generation with AGrid partitioning, we consider the following percentage splits: 12.5--12.5--75; 20--20--60; 25--25--50; 33--33--33; and 40--40--20.
We find that $\epsilon_1 = \epsilon_2 = 0.4\epsilon$ and $\epsilon_3 = 0.2\epsilon$ gives good results.  
For cluster-based partitioning \textit{without} KDE, we find that $\epsilon_1 = 2\epsilon_2$ is the best setting.
For cluster-based partitioning \textit{with} KDE, we find that setting $\epsilon_3 = 0.25\epsilon$ is best, which leaves $\epsilon_1 = 0.25\epsilon$ and $\epsilon_2 = 0.5\epsilon$.
As noted previously, cluster-based partitioning generally leads to flatter kernels as regions tend to be larger, and so a slightly higher $\epsilon_3$ value helps to keep $h$ at a value that prevents the kernel from becoming too flat.
For Road, equal division of $\epsilon$ balances noise added to edges with noise added to the micro-histograms.
We use these allocations as the default settings throughout. 

\para{Number of Clusters}
When Clust-KDE is used, NCE values decrease as the number of initial clusters increases.  This is intuitive as regions are smaller, which allows the kernel density estimate to be better tailored to the characteristics of the regions.

\subsubsection{Real World Considerations}
\label{sss:real-world-considerations}
We next evaluate how well our methods model characteristics of real world data, which is often messy and can exhibit high non-uniformity or skew. 

\para{Road Network Alignment}
For Road, we assume that data points are well-aligned with the underlying road network.  However, this is not always the case with real datasets, and there can be high error when map-matching raw data points to edges in the road network. This may be due to GPS sampling errors, map projection errors, and multi-lane roads being modeled as single lines of zero width.  

Whereas we use `uncorrected' data in the main experiments, we now perform experiments where we use the map-matched data as the input datasets (i.e., $d(p, e_p) = 0$). 
In this new setting, we find that Road is far superior to the other methods, which perform up to 18\% worse.  Hence, when the data is corrected, Road is up to 10\%, 10\%, and 120\% more accurate than UGrid-KDE, AGrid-KDE, and Clust-KDE, respectively.

\para{Uneven Population Densities}
Population density in cities is rarely uniform, either across an area, or along individual roads.
In the urban centers, point density may be somewhat uniform along edges, while rural and suburban areas may experience more varied densities. 
To examine how our methods are affected by uneven densities, we create a dataset focused on a larger area of Beijing, which includes more suburban areas.
We set the expanded bounds of the studied region to the bounding box between (116.33, 39.97) and (116.48, 39.85).  
In the new road network, $|E| = 13{,}862$ and so we set $N = 20|E| = 277{,}240$.  
We find that both UGrid-KDE and Road are relatively robust, but AGrid-KDE and Clust-KDE perform worse.

\subsection{Range and Hotspot Queries}
\label{ss:range-hotspot}

\subsubsection{Range Queries}
\label{sss:range}
\begin{figure}[t]
    \centering
    \begin{subfigure}[b]{\columnwidth}
        \centering
        \includegraphics[width=6cm]{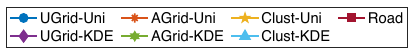}
    \end{subfigure}
    \\
    \begin{subfigure}[b]{0.325\columnwidth}
        \centering
        \includegraphics[height=4.5cm]{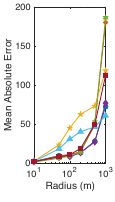}
        \caption{Beijing}
        \label{fig:beijing-range}
    \end{subfigure}
   \hfill
    \begin{subfigure}[b]{0.325\columnwidth}
        \centering
        \includegraphics[height=4.5cm]{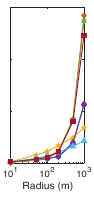}
        \caption{Porto}
        \label{fig:porto-range}
    \end{subfigure}
    \hfill
    \begin{subfigure}[b]{0.325\columnwidth}
        \centering
        \includegraphics[height=4.5cm]{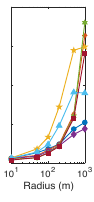}
        \caption{New York}
        \label{fig:nyc-range}
    \end{subfigure}
    \caption{Variation in MAE of range queries as \textit{r} varies}
    \label{fig:range}
\end{figure}

Range queries are important in location analytics to quickly assess how many customers are potentially available to a business, measure accessibility to key services within a certain time, etc. 
To assess this, we specify a set, $\mathcal{L}$, of 100 arbitrary locations in each city (selected from the set of nodes in each city's road network), and specify a circular region defined by the radius, $r$.  
To quantify the error, we use mean absolute error (MAE), in which $c_{l}^{\real}$ and $c_{l}^{\synth}$ respectively denote the number of real and synthetic points within $r$ meters of location $l$, and:
\begin{equation}
    MAE = \frac{1}{|\mathcal{L}|} \sum_{l \in \mathcal{L}} \left|c_{l}^{\real} - c_{l}^{\synth}\right|
    \label{eq:mae}
\end{equation}
Figure \ref{fig:range} shows how the radius of the range query influences the error, for each method and city. 
For small $r$, all partitioning-based methods outperform their respective baselines.
Interestingly, although Clust-KDE is generally less competitive, it performs better in the less-ordered Porto.
Road is a viable alternative when $r$ is small; although, as $r$ increases, its error increases rapidly.
Likewise, AGrid methods perform notably worse for large $r$ values.
However, when one considers the error in relation to the dataset size, as well as the proportion of the query range to the entire dataset domain, this behavior is acceptable.
Despite this, UGrid methods offer strong alternatives, depending on the degree of road network alignment.

\subsubsection{Hotspot Queries}
\label{sss:hotspot}
Hotspot queries are also fundamental in location analytics for businesses to identify popular regions for advertising, for city agencies to help manage congestion and traffic flow, etc.
Here, we obtain kernel density estimates for the real and synthetic datasets, at varying granularities.  
We use a Gaussian kernel over a $g \times g$ uniform grid, where $g$ denotes the grid granularity; we use granularities: $g = \{2^6, 2^7, 2^8, 2^9, 2^{10}\}$.  Note that our kernel function can be non-private (i.e., the kernel is tuned to the data) here as we are simply assessing the utility of the output data. 
We define hotspots to be locations with a density greater than the 95\textsuperscript{th} percentile.  
To assess query response similarity between the two datasets, we use the S\o{}rensen-Dice coefficient (SDC), defined as:
\begin{equation}
    SDC = \frac{2\big|\mathcal{H}^{\real} \bigcap \mathcal{H}^{\synth}\big|}{\big|\mathcal{H}^{\real}\big| +  \big|\mathcal{H}^{\synth}\big|}
    \label{eq:sorensen-dice}
\end{equation}
where $\mathcal{H}$ is the set of hotspots.

\begin{figure}[t]
    \centering
    \begin{subfigure}[b]{\columnwidth}
        \centering
        \includegraphics[width=6cm]{Figures/legend7.pdf}
    \end{subfigure}
    \\
    \begin{subfigure}[b]{0.32\columnwidth}
        \centering
        \includegraphics[height=4.5cm]{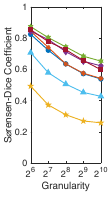}
        \caption{Beijing}
        \label{fig:beijing-hotspot}
    \end{subfigure}
   \hfill
    \begin{subfigure}[b]{0.32\columnwidth}
        \centering
        \includegraphics[height=4.5cm]{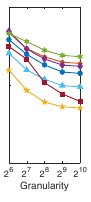}
        \caption{Porto}
        \label{fig:porto-hotspot}
    \end{subfigure}
    \hfill
    \begin{subfigure}[b]{0.32\columnwidth}
        \centering
        \includegraphics[height=4.5cm]{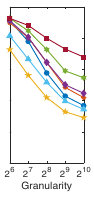}
        \caption{New York}
        \label{fig:nyc-hotspot}
    \end{subfigure}
    \caption{Variation in SDC as the hotspot granularity varies}
    \label{fig:hotspots}
\end{figure}

Figure \ref{fig:hotspots} shows similarity decrease as granularity increases, as the kernel density estimates are more sensitive to small changes in the location of individual points.  All partitioning-based methods outperform their respective baselines, and Road performs especially well when the original data is well-aligned with the road network (e.g., New York, Figure \ref{fig:nyc-hotspot}). However, Road performs less well with dense road networks or poorly aligned data (e.g., Porto, Figure \ref{fig:porto-hotspot}).  
Conversely, grid-based methods perform better in less-structured environments, but perform worse when data is well-aligned with the roads.

\subsection{Facility Location Queries}
\label{ss:fl}
Facility location is a common analytics task for which individual location data is necessary and is one possible application for our methods.  
Given a set $\mathcal{F}$ of existing facilities and a set $\mathcal{C}$ of candidate facilities, a facility location query (FLQ) aims to find the best $B$ locations that satisfy a stated objective function.  
We consider the two most common FLQ variants. 
In the \textsc{Max-Inf} case, we seek to identify the most influential candidate facilities, where influence is commonly defined as the total number of customers that the facilities attract.  
In the \textsc{Min-Dist} case, we find the facilities that minimize the total distance between customers and the facilities.  

\subsubsection{Outline}
\label{sss:fl-outline}
Consider the case where a food stand company wishes to locate a number of outlets in the center of Beijing.  
We intuit that a lot of business would be generated if the outlets were located at the intersections of busy roads and so we use the location set, $\mathcal{L}$, (from Section \ref{sss:range}) where each $l \in \mathcal{L}$ now represents a candidate facility.
For the real dataset, we use those from Section \ref{s:expts} and, for the synthetic datasets, we use those generated in Section \ref{s:expts} under the default conditions.
We assume that there are no existing facilities currently in the city (i.e., $\mathcal{F}$ = \O).  
We also define $\mathcal{B^{\real}}$ and $\mathcal{B^{\synth}}$ to be the sets of selected facilities when the real and synthetic datasets are used, respectively.  
We use the SDC to assess accuracy of FLQs when using synthetic data.  In this setting, the SDC will capture the extent to which synthetic data identifies the same top-$B$ facilities as the real data.
We use $\mathcal{B^{\real}}$ and $\mathcal{B^{\synth}}$ in place of $\mathcal{H^{\real}}$ and $\mathcal{H^{\synth}}$ from Equation \ref{eq:sorensen-dice}.

\subsubsection{Results}
\label{sss:fl-results}
Table~\ref{tab:fl} shows the SDC values (when $B = 20$) for both FLQs.  
We see that, irrespective of the data generation method,  both variants of FLQs are answered almost identically compared to when the real data is used.
This is because the optimal locations are quite robust to the noise added to achieve DP. 
The SDC values indicate that at least 19 of `true' top 20 candidate facilities are selected when using the synthetic data, which further highlights its suitability for answering FLQs.
We also explore the effect that changing $B$ has on the SDC values.  
Our methods are robust and perform equally well for all values of $B$.
In particular, they produce optimal results to the \textsc{Min-Dist} FLQ for all values of $B$.

There may be some cases in which using synthetic data does not obtain similar results to FLQs.  
For example, when candidate facilities are close to each other, customers may be assigned to different facilities if their location is perturbed a little.  
Another example is in the capacitated facility location problem (when capacity constraints are strict) when  `additional' customers generated through additive noise cannot be accommodated at their nearest facility.
However, overall our methods generate synthetic data that exhibit high levels of accuracy for FLQs compared to using real data.
In reality, this means that researchers and companies do not need use real data for facility location.
Instead, private synthetic data can be used without compromising on the accuracy of the facility location analysis.

\begin{table}[tb]
    \centering
    \caption{S\o{}rensen-Dice Coefficients ($\bm{B=20}$) for FLQs}
    \label{tab:fl}
    \begin{tabular}{cccccccc}
        \toprule
        \textbf{Data Gen.} & \multicolumn{2}{c}{\textbf{UGrid}} & \multicolumn{2}{c}{\textbf{AGrid}} & \multicolumn{2}{c}{\textbf{Clust}} & \multirow{2}{*}{\textbf{Road}}\\
        \textbf{Method} & \textbf{Uni} & \textbf{KDE} & \textbf{Uni} & \textbf{KDE} & \textbf{Uni} & \textbf{KDE} & \\
        \midrule 
        \textsc{Max-Inf} & 1.00 & 1.00 & 1.00 & 1.00 & 0.95 & 0.95 & 0.95 \\ 
        \textsc{Min-Dist} & 1.00 & 1.00 & 1.00 & 1.00 & 1.00 & 1.00 & 1.00 \\ 
     \bottomrule 
    \end{tabular}
\end{table}

\subsection{Discussion}
\label{s:discussion}
While both partitioning-based and road network-based approaches are effective in practice, different methods are more appropriate for different circumstances.  We summarize our findings here.

All methods scale well in accuracy terms.  In particular, Road accommodates large datasets easily, and the error decreases with input size. Hence, Road should be the default data generation method, especially when the raw data is well-aligned with the road network.
Where road network data is unavailable or the data is poorly aligned with the road network, partitioning-based approaches should be considered.
UGrid-KDE and AGrid-KDE are generally comparable, although AGrid methods are particularly strong in more structured environments.
For very large datasets, the difference in runtime costs between  clustering- and grid-based methods is larger (cf. Equation \ref{eq:min-grid-size} and Figure \ref{fig:pb-time} -- $N$ and $\epsilon$ have similar effects on runtime), and so clustering-based methods should be considered in this case. 

For facility location analytics tasks, all methods perform very well and all methods can be recommended as a general purpose solution. 
For range queries, all methods are highly effective especially when the range query radius is small.  
If the range query radius is large, UGrid approaches are recommended (with consideration of the degree of network alignment).
For hotspot queries, we advise using Road for datasets that are well-aligned with the road network, which is the case for most applications.
UGrid-KDE and AGrid-KDE are more effective when the datasets are less well-aligned, or when the road network is less well-structured. 

\section{Conclusion}
\label{s:conc}
In this paper, we introduced novel approaches for generating synthetic location data that satisfy the requirements of $\epsilon$-differential privacy.  
The proposed methods ensure that the generated data preserves the underlying characteristics of the real data, while ensuring that the existence and location of all individuals remains private. An extensive series of experiments confirms that the generated synthetic data has a high degree of similarity with the real data upon which it is based. We achieve further practical utility by incorporating public knowledge, such as road networks, coastlines, and rivers, within our methods. We have also applied our data generation methods to a range of location analytics queries and shown that the synthetic data obtains excellent results compared to the results obtained with real data.  These strong results pave the way for everyday practical use of differential privacy in the real world.

\balance

\begin{acks}
The authors thank Emre Yilmaz for his early work on this paper.
This work is supported by the UK Engineering and Physical Sciences Research Council (EPSRC) under Grant No. EP/L016400/1, European Research Council under Grant No. ERC-2014-CoG 647557, and by The Alan Turing Institute under EPSRC Grant No. EP/N510129/1.

\end{acks}

\bibliographystyle{ACM-Reference-Format}
\bibliography{99-bib.bib}

\end{document}